\documentclass[aps, prb, amsmath, amssymb, reprint]{revtex4-1}

\usepackage[utf8]{inputenc} 

\usepackage{graphicx}
\usepackage{color}
\usepackage{cases,array}
\usepackage{bm}

\usepackage[english]{babel}

\usepackage{xcolor}

\usepackage{scrextend}

\usepackage{hyperref}

\hypersetup{
	colorlinks,
	linkcolor={blue},
	citecolor={blue},
	urlcolor={blue}
}

\usepackage{commath}

\newcommand{\me}{\mu_\text{eff}}

\newcommand{\mv}[1]{\mathbf{#1}}

\newcommand{\muv}[1]{\hat{\mv{#1}}}

\newcommand{\mvd}[1]{\dot{\mv{#1}}}

\newcommand{\SIN}[1]{\sin\left(#1\right)}
\newcommand{\COS}[1]{\cos\left(#1\right)}

\newcommand{\ABS}[1]{\left| #1 \right|}
\newcommand{\MEAN}[1]{\left\langle #1 \right\rangle}
\newcommand{\ABSMEAN}[1]{\ABS{\MEAN{#1}}}

\newcommand{\G}[1]{\mv{G}_{#1}}
\newcommand{\D}[1]{\overset{\leftrightarrow}{\mv{D}}_{#1}}

\newcommand{\mytab}[2]{
	\renewcommand{\arraystretch}{1.0}
	\begin{tabular}{@{}c@{}} #1\\ #2 \end{tabular}
}

\usepackage{empheq}

\begin{document}
	
\title{Optimizing magneto-dipolar interactions for synchronizing vortex based spin-torque nano-oscillators}

\author{F. Abreu Araujo}
\email{abreuaraujo.flavio@gmail.com}
\affiliation{Institute of Condensed Matter and Nanosciences, Universit\'{e} catholique de Louvain, Place Croix du Sud 1, 1348 Louvain-la-Neuve, Belgium}

\author{A.D. Belanovsky}
\affiliation{A. M. Prokhorov General Physics Institute, RAS, Vavilova, 38, 119991 Moscow, Russia}
\affiliation{Moscow Institute of Physics and Technology, Institutskiy per. 9, 141700 Dolgoprudny, Russia}

\author{N. Locatelli}
\altaffiliation[Present address: ]{Institut d’Electronique Fondamentale, UMR CNRS 8622, Univ. Paris-Sud, 91405 Orsay, France}
\affiliation{Unit\'e Mixte de Physique CNRS/Thales, 1 ave A. Fresnel, 91767 Palaiseau, and Universit\'{e} Paris-Sud, 91405 Orsay, France}

\author{R. Lebrun}
\affiliation{Unit\'e Mixte de Physique CNRS/Thales, 1 ave A. Fresnel, 91767 Palaiseau, and Universit\'{e} Paris-Sud, 91405 Orsay, France}

\author{G. de Loubens}
\affiliation{Service de Physique de l'\'{E}tat Condens\'{e} (CNRS URA 2464), CEA Saclay, 91191 Gif-sur-Yvette, France}

\author{O. Klein}
\altaffiliation[Present address: ]{SPINTEC, UMR CEA/CNRS/UJF-Grenoble 1/Grenoble-INP, INAC, 38054 Grenoble, France}
\affiliation{Service de Physique de l'\'{E}tat Condens\'{e} (CNRS URA 2464), CEA Saclay, 91191 Gif-sur-Yvette, France}

\author{P.N. Skirdkov}
\affiliation{A. M. Prokhorov General Physics Institute, RAS, Vavilova, 38, 119991 Moscow, Russia}
\affiliation{Moscow Institute of Physics and Technology, Institutskiy per. 9, 141700 Dolgoprudny, Russia}

\author{K.A. Zvezdin}
\altaffiliation[Also at: ]{Istituto P.M. srl, via Grassi, 4, 10138, Torino, Italy}
\affiliation{A. M. Prokhorov General Physics Institute, RAS, Vavilova, 38, 119991 Moscow, Russia}
\affiliation{Moscow Institute of Physics and Technology, Institutskiy per. 9, 141700 Dolgoprudny, Russia}

\author{J. Grollier}
\affiliation{Unit\'e Mixte de Physique CNRS/Thales, 1 ave A. Fresnel, 91767 Palaiseau, and Universit\'{e} Paris-Sud, 91405 Orsay, France}

\author{A.K. Zvezdin}
\affiliation{A. M. Prokhorov General Physics Institute, RAS, Vavilova, 38, 119991 Moscow, Russia}
\affiliation{Moscow Institute of Physics and Technology, Institutskiy per. 9, 141700 Dolgoprudny, Russia}

\author{V. Cros}
\affiliation{Unit\'e Mixte de Physique CNRS/Thales, 1 ave A. Fresnel, 91767 Palaiseau, and Universit\'{e} Paris-Sud, 91405 Orsay, France}

\begin{abstract}
We report on a theoretical study about the magneto-dipolar coupling and synchronization between two vortex-based spin-torque nano-oscillators. In this work we study the dependence of the coupling efficiency on the relative magnetization parameters of the vortices in the system. For that purpose, we combine micromagnetic simulations, Thiele equation approach, and analytical macro-dipole approximation model to identify the optimized configuration for achieving phase-locking between neighboring oscillators. Notably, we compare vortices configurations with parallel (P) polarities and with opposite (AP) polarities. We demonstrate that the AP core configuration exhibits a coupling strength about three times larger than in the P core configuration.
\end{abstract}

\maketitle

\section{Introduction}

In the last decade great attention has been drawn to the phase-locking phenomena of spin-torque nano-oscillators (STNOs) \cite{Slavin2005, Slavin2006, Grollier2006, Rezende2007, Georges2008, Ruotolo2009, Bonin2009, Chen2009, Slavin2009_2, Tiberkevich2009, Zhou2009, Tabor2010, Urazhdin2010, Chen2011, Li2011, dAquino2012, Finocchio2012, Li2012, Nakada2012, Jain2012, Belanovsky2012, Belanovsky2013, Jain2014}. STNOs are anticipated to be promising devices for sub-micron scale microwave synthesizers because of their high emission frequency tunability\cite{Kiselev2003, Kaka2005, Russek2010}. However, an important issue of such devices regarding their practical realization is their low output oscillation power and low spectral stability. A possible solution to these issues could be the synchronization of a few STNOs~\cite{Ruotolo2009, Russek2010, Li2010, Nakada2012, Belanovsky2012, Belanovsky2013}. Synchronization between multiple auto-oscillators can also be useful in the framework of developing associative memories architectures~\cite{Csaba2012, Csaba2012_2, Roska2012, Shibata2012}. Previous studies reported on synchronization of STNOs interacting with others via spin waves\cite{Russek2010, Kaka2005, Mancoff2005}, exchange coupling\cite{Ruotolo2009}, electric currents\cite{Li2010, Grollier2006, Georges2008_2}, noisy current injection\cite{Nakada2012}, or via magneto-dipolar interaction\cite{Shibata2003, Vogel2010, Awad2010, Barman2010, Jung2010, Jung2011, Belanovsky2012, Belanovsky2013, Berkov2013}.

Among the various synchronization mechanisms, magneto-dipolar coupling is inherent and efficient as emphasized in our previous works \cite{Belanovsky2012, Belanovsky2013} but also in refs \cite{Shibata2003, Vogel2010, Awad2010, Barman2010, Jung2010, Jung2011, Berkov2013}. In the present study, we focus on the magnetodipolar interaction between two vortex based STNOs.



Single magnetic vortices in cylindrical dots are characterized by two topological parameters\cite{Shibata2003}. Chirality ($C$) determines the curling direction of the in-plane magnetization, such that $C = +1$ (resp. $C = -1$) stands for counter-clockwise (resp. clockwise) direction. The orientation of the vortex core magnetization is described by its polarity ($P$), which takes a value of $P = +1$ (resp. $P = -1$) for core magnetization aligned (resp. anti-aligned) to the out-of-plane ($\hat{z}$) axis. The relative configuration of two interacting vortices can then take four non-equivalent states, with identical/opposite chiralities and identical/opposite polarities.

In a previous work \cite{Belanovsky2012, Belanovsky2013}, we studied the capability of two vortex-based STNOs to synchronize through dipolar coupling. In this first approach, we have only considered the case of two vortices with identical polarities and chiralities, and already demonstrated the possibility to observe synchronization. In this new study report, we show that changing the relative polarity and chirality parameters will strongly modify the interaction between the auto-oscillators and may strongly modify the efficiency of synchronization. We conduct a numerical study in which we investigate the synchronization properties for selected combinations of vortex parameters, aiming at sorting the best combinations of the ($C$, $P$) parameters to achieve synchronization. We also consider two different electrical connections for the current injection i.e, parallel and series connections, corresponding respectively to current flowing in the same or opposite direction in the two STVOs.

\section{Presentation of the system}

\begin{figure}[!ht]
	\includegraphics[scale=1]{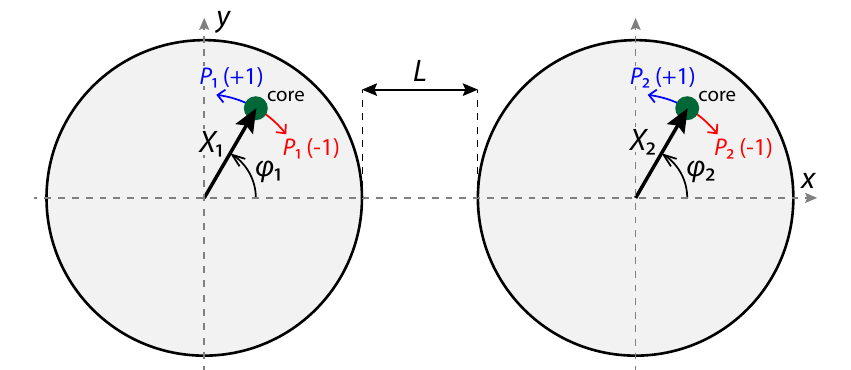}
	\caption{(Color online) Schematic illustration of the studied system composed of two magnetic dots, each in a magnetic vortex configuration. Vortices cores are shown by small green disks. The vortex core positions are given in polar coordinates, i.e. $(X_1,\varphi_1)$ and $(X_2,\varphi_2)$, respectively. The blue (resp. red) arrows show the core up (resp. down) gyrotropic motion sense.}
	\label{fig:studiedsyst}
\end{figure}

The studied system consists of two circular nanopillars with identical diameters $2R = 200$ nm, separated by an interdot distance $L$ (see Fig. \ref{fig:studiedsyst}). Each incorporates a Permalloy free magnetic layer ($M_\text{s} = 800$ emu/cm$^3$, $A = 1.3 \times 10^{-6}$ erg/cm, $\alpha = 0.01$) with thickness $h=10$ nm, separated by an intermediate layer (non-magnetic metal or tunnel barrier) from a polarizing layer with perpendicular magnetization. Considering their dimensions, each free layer has a magnetic vortex as its remnant magnetic configuration. The vortices parameters will be referred as $P_{1,2}$ and $C_{1,2}$ for 1st and 2nd pillar. The polarizing layers, whose magnetizations are identical and oriented along $\hat{z}$, will be considered in simulations only through the corresponding current spin polarization $p_{z1}=p_{z2}=p_z=+0.2$. 
The gyrotropic motion of a vortex core can be driven by spin transfer torque action, by flowing current above a threshold amplitude through each pillar ; In our case, a current density $J = 7 \times 10^6$ A/cm$^2$ ($I_\text{DC}=2.2$ mA). Yet, the current sign in each pillar has to be chosen so that $I_i P_i p_z < 0$ to ensure self-sustained oscillations~\cite{Dussaux2010, Khvalkovskiy2009}. The core polarity of each vortex then defines its gyration direction\cite{Guslienko2002}  (see Fig. \ref{fig:studiedsyst}). Indeed, when $P_i = +1$ (resp. $P_i = -1$) the vortex core circular motion is counterclockwise (resp. clockwise).

\section{Possible configurations}

We then consider six possible configurations for which self-sustained oscillations are achieved in both pillars, described in Table~\ref{tab:tab1}. Note that Parallel cores (Pc) configurations correspond to vortices moving in the same direction, whereas Anti-Parallel cores (APc) configurations correspond to vortices moving in the opposite directions.

\begin{table}[!ht]
	\centering
	\begin{ruledtabular}
		\begin{tabular}{ccccccccc}
			& \multicolumn{4}{c}{Left Dot} & \multicolumn{4}{c}{Right Dot}\\
			\cline{2-5}                               \cline{6-9}
			config.  & $C_1$ & $P_1$ & $J_1$ & $p_{z1}$ & $C_2$ & $P_2$ & $J_2$ & $p_{z2}$\\
			\hline
			Pc1  & $-1$ & $-1$ & $+$ & $+0.2$ & $-1$ & $-1$ & $+$ & $+0.2$\\
			Pc2  & $+1$ & $-1$ & $+$ & $+0.2$ & $+1$ & $-1$ & $+$ & $+0.2$\\
			Pc3  & $-1$ & $-1$ & $+$ & $+0.2$ & $+1$ & $-1$ & $+$ & $+0.2$\\
			APc1 & $-1$ & $-1$ & $+$ & $+0.2$ & $+1$ & $+1$ & $-$ & $+0.2$\\
			APc2 & $+1$ & $-1$ & $+$ & $+0.2$ & $-1$ & $+1$ & $-$ & $+0.2$\\
			APc3 & $-1$ & $-1$ & $+$ & $+0.2$ & $-1$ & $+1$ & $-$ & $+0.2$\\
		\end{tabular}
	\end{ruledtabular}
	\caption{\label{tab:tab1}Studied configurations with their respective signs of the vortex parameters ($C_i$, $P_i$), current density ($J_i$), and current spin polarization ($p_{zi}$).}
\end{table}

\begin{figure}[!ht]
	\includegraphics[scale=1]{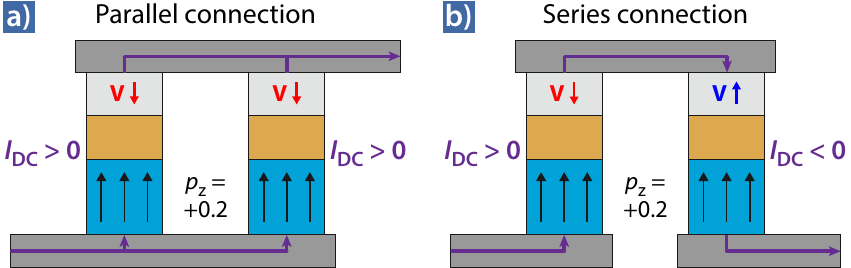}
	\caption{(Color online) a) (resp. b)) Illustration of the DC supplied current for the P (resp. AP) configurations showing the parallel (resp. series) connection.}
	\label{fig:connection}
\end{figure}

Considering the different configurations displayed in Tab. \ref{tab:tab1}, the electrical connection must be adapted according to the relative vortex core polarities in order to fulfill the condition $I_i P_i p_z < 0$ to ensure self-sustained oscillations~\cite{Dussaux2010, Khvalkovskiy2009}. As a consequence, the Pc configurations must be alimented using the parallel connection to ensure same current sign in both pillars (see Fig. \ref{fig:connection}a). On the contrary, the APc configurations have to be supplied with a series connection to ensure opposite current signs (see Fig. \ref{fig:connection}b).

\section{\label{sec:macrodip}Macro-dipole analytical model}

To get some insights for the origin of the dependence in effective coupling with vortices configuration, we concentrate in this section on an analytical model based on a macro-dipole approximation. The dipolar energy ($W_\text{int}$) between two magnetic dipoles $\mv{\mu}_1$ and $\mv{\mu}_2$ is then given by the following equation (in CGS units):
\begin{equation}
W_\text{int} = -\frac{(3(\mv{\mu}_1\cdot\mv{e}_{12})(\mv{\mu}_2\cdot\mv{e}_{12})-\mv{\mu}_1\cdot\mv{\mu}_2)}{\Vert\mv{D}_{12}\Vert^3},
\label{dipen}
\end{equation}
where $\mv{D}_{12}$ is the vector between the two dipoles and $\mv{e}_{12}$ is a unit vector parallel to $\mv{D}_{12}$.

Considering two planar dipoles induced by the off-centered vortices in the framework of the two vortex ansatz \footnote{In the two vortex ansatz model the volume averaged magnetization of the shifted vortex is proportional to its displacement {$\mv{\mu}/V = \MEAN{ \mv{M}(\mv{X}(t)) }_V = -\xi C M_\text{s} / R \left[ \muv{z} \times \mv{X}(t) \right]$}, where {$V = \pi R^2 h$}, $\mv{X}(t) = X \left[ \COS{\varphi(t)}, \SIN{\varphi(t)} \right]$, and {$\xi = 2/3$} for this model \cite{Guslienko2002, Guslienko2006_APL}} (TVA): $\mv{\mu}_1 = \sigma C_1 X_1 (-\SIN{\varphi_1}, \COS{\varphi_1})$ and $\mv{\mu}_2 = \sigma C_2 X_2 (-\SIN{\varphi_2}, \COS{\varphi_2})$, where $\sigma = \xi M_\text{s} V / R$, $\xi = 2/3$, $V = \pi R^2 h$ . For $\mv{D}_{12} = (d,0)$, where $d=2R+L$ is the inter-dipole distance along $x$-axis, and using equation (\ref{dipen}) one obtains:
\begin{equation}
W_\text{int} = - C_1 C_2 \frac{\sigma^2}{2 d^3} X_1 X_2 (\cos(\varphi_1-\varphi_2) - 3\cos(\varphi_1+\varphi_2))
\label{dipen2}
\end{equation}
\noindent where $\dot{\varphi_i} = P_i \omega_i$.  

To illustrate the different situations, we consider synchronized oscillations in the two relative polarities configurations. For two vortices with same core polarity (Pc), gyrating in identical directions at the same frequency $\varphi_1-\varphi_2 \approx 0$ and $\varphi_1+\varphi_2 \approx 2\omega_0$, so that equation (\ref{dipen2}) gives:
\begin{equation}
W_\text{int}^\text{Pc} = - C_1 C_2 \frac{\sigma^2}{2 d^3} X_1 X_2 (1 - 3\COS{2\omega_0 t})
\label{eq:dipenIP}
\end{equation}

\noindent In contrast, for vortices with opposite polarities (APc), gyrating in opposite directions, $\varphi_1+\varphi_2 \approx 0 $ and $\varphi_1-\varphi_2 \approx 2\omega_0$, so that one obtains:
\begin{equation}
W_\text{int}^\text{APc} = -C_1 C_2 \frac{\sigma^2}{2 d^3} X_1 X_2 (\COS{2\omega_0 t} - 3)
\label{eq:dipenOOP}
\end{equation}

\begin{figure}[!ht]
	\includegraphics[scale=1]{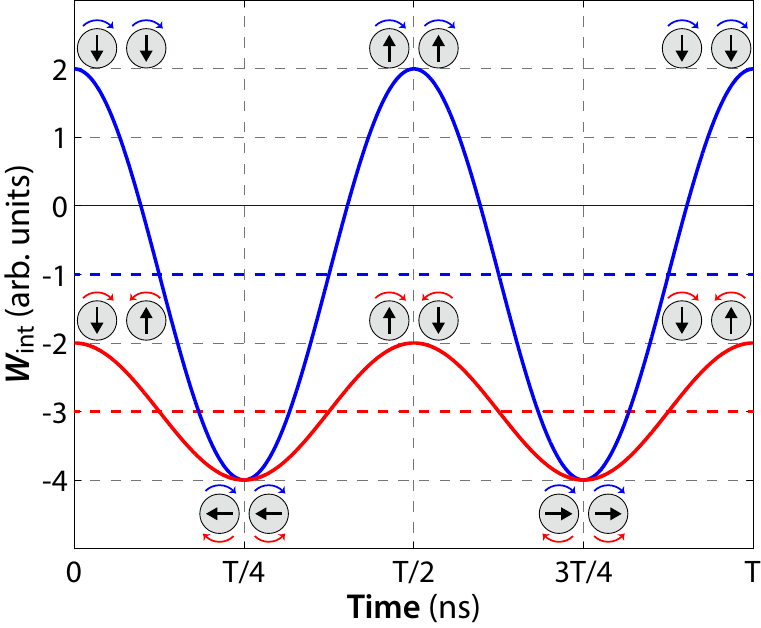}
	\caption{(Color online) a) Dipolar energy ($W_\text{int}$) evolution of two interacting vortices modelled as macro-dipoles and oscillating at the same frequency. The blue curve corresponds to the identical polarities (Pc) case and the red curve corresponds to the opposite polarities (APc) case. The dashed colored lines represent the corresponding mean value of the coupling energies $\MEAN{ W_\text{int} }$.}
	\label{fig:macrodip}
\end{figure}

Equations (\ref{eq:dipenIP}) and (\ref{eq:dipenOOP}) show that for a given vortex gyration frequency $\omega_0$ the coupling energy $W_\text{int}$ oscillates at twice the frequency ($2\omega_0$). In the Pc case (see blue curve in Fig. \ref{fig:macrodip}) it oscillates with a large amplitude and a small mean value, whereas in the APc case (see red curve in Fig. \ref{fig:macrodip}) it oscillates with a large amplitude and a smaller mean value.

When not synchronized, the two vortices will feel two oscillating components of the magneto-dipolar interaction i.e., one at low frequency and one at high frequency. The latter one will average out and have negligible influence on the phase locking features, while the low frequency term will be responsible for the synchronization phenomenon. The effective coupling coefficient $\me$ can be identified writing $\MEAN{ W_\text{int} } = \me C_1 C_2 X_1 X_2$ for the mean coupling energy and gives the following results for the Pc and APc relative vortex core polarity configurations:
\begin{empheq}[left=\empheqlbrace]{align*}
\me^\text{Pc}  &= - \frac{\pi^2 \xi^2 R^2 h^2}{2 d^3}\\
\me^\text{APc} &= 3 \frac{\pi^2 \xi^2 R^2 h^2}{2 d^3}
\end{empheq}

\begin{figure}[!ht]
	\includegraphics[scale=1]{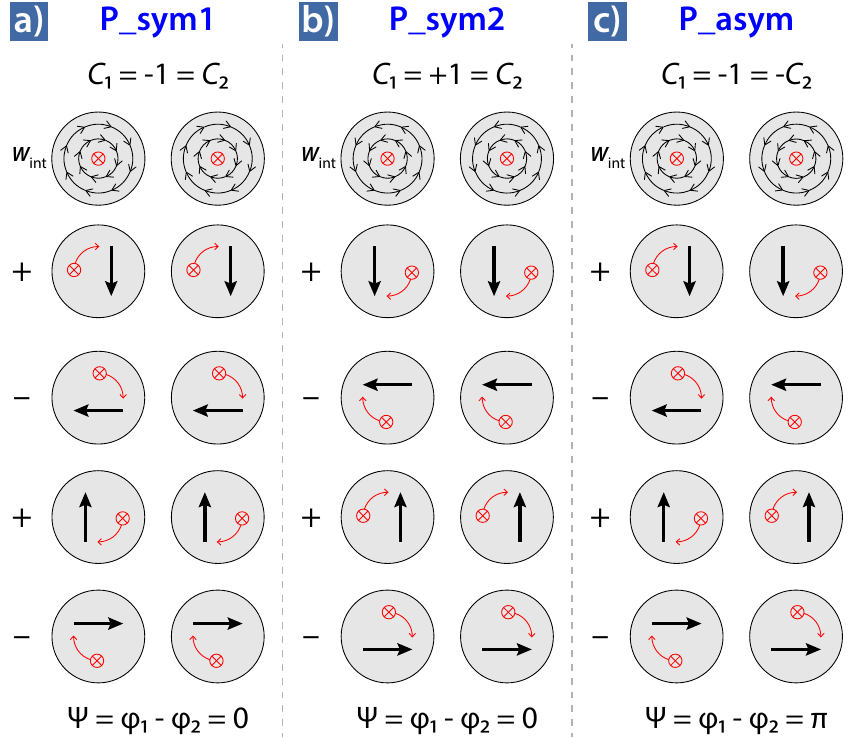}
	\caption{(Color online) Schematic illustration of the coupled dynamics for the Pc configurations, i.e. where $P_1 = P_2$.}
	\label{fig:Pcases}
\end{figure}

\begin{figure}[!ht]
	\includegraphics[scale=1]{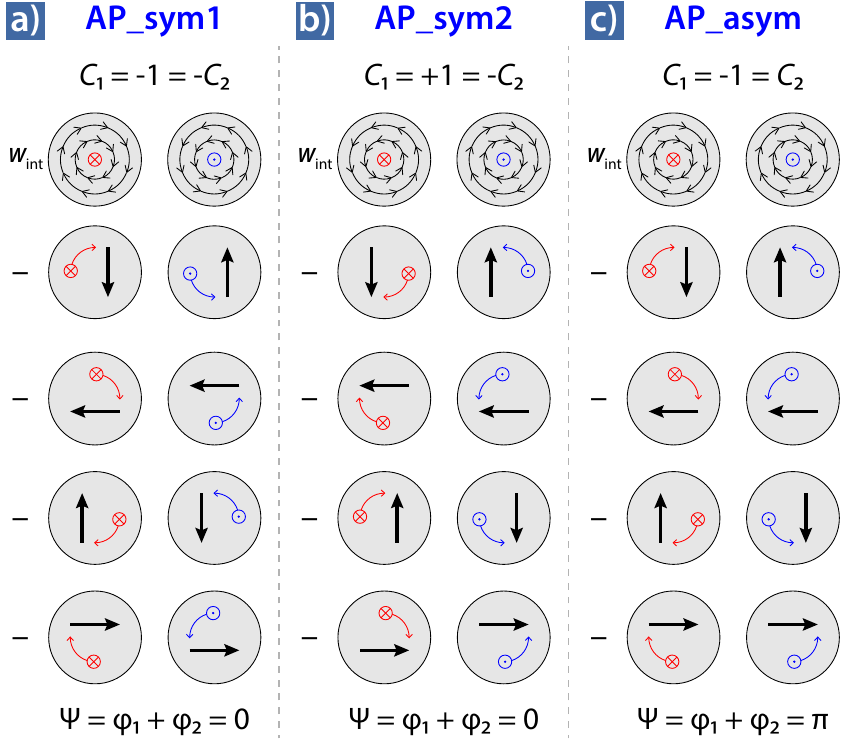}
	\caption{(Color online) Schematic illustration of the coupled dynamics for the APc configurations, i.e. where $P_1 = -P_2$.}
	\label{fig:APcases}
\end{figure}

Synchronized states correspond to a minimization of the average interaction energy. As illustrated here, relative polarities and chiralities signs influences the sign of $W_\text{int}$. As a consequence, these relative parameters also define the phase relationship achieved when synchronization occurs. The later considerations are illustrated in figures \ref{fig:Pcases} and \ref{fig:APcases}.

From this study, we then conclude that the effective coupling coefficient is predicted to be three times stronger when polarities are opposite (APc) than when polarities are identical (Pc). Concurrently, the high frequency oscillation of interaction energy is three times larger in Pc polarities configuration as compared to APc case. While this indicates that APc is the optimal configuration for synchronization, we must note that this second contribution may affect the locking phenomenon. 

\section{\label{sec:thiele}Thiele analytical approach}


The spin transfer induced gyrotropic vortex dynamics can be described by the Thiele equation \cite{Thiele1973, Khvalkovskiy2009, Gaididei2010, Belanovsky2012}:
\begin{align}
\begin{array}{c}
\displaystyle P_i \G{i} \times \mvd{X}_i + \D{i} \cdot \mvd{X}_i - k_i (X_i,C_i,J_i) \mv{X}_i\\[10pt]
\displaystyle -\mv{F}^\text{STT}_i(p_{zi},J_{i},P_{i}) - \mv{F}_\text{int}(\mv{X}_j, P_i P_j, C_i C_j) = \mv{0}
\end{array}
\label{eq:Thiele}
\end{align}
where $\G{i} = -G \muv{z}$ is the gyrovector with $G = 2 \pi M_\text{s} h/\gamma$ and $\D{i} = \alpha \eta_i G$ is the damping coefficient  with $\eta_i = 0.5 \ln \left( R_i / (2L_\text{ex}) \right) + 3/8$,\cite{Khvalkovskiy2009} where $L_\text{ex}$ is the exchange length. For each pillar, the vortex frequency\cite{Guslienko2006, Ivanov2007} is given by the ratio between the confinement coefficient $k_i$ and the gyrovector $\omega_{0\mid i}=k_{i}/G$ with:
\begin{equation}
k_i (X_i, C_i, J_i) = k^\text{ms}_i + k^\text{Oe}_i C_i J_i + (k^{\prime\text{ms}}_i + k^{\prime\text{Oe}}_i C_i J_i) \left( \frac{\mv{X}_i^2}{R_i^2} \right)
\label{eq:k}
\end{equation}
\noindent where $k^\text{ms}_i$ and $k^{\prime\text{ms}}_i$ (resp. $k^\text{Oe}_i$ and $k^{\prime\text{Oe}}_i$) corresponds to the magnetostatic (resp. Oersted field) contribution. The Oersted contribution will increase the vortex core gyration frequency if the vortex chirality is along the same direction as the Oersted field ($ C_i J_i > 0$), and respectively decrease the frequency otherwise\cite{Dussaux2012}. Gyration amplitudes will also be affected by such interaction with Oersted field. 
To maximize the symmetry of the system and avoid that the Oersted contribution brings an offset between the two STNOs' frequencies, we find that the condition $C_1 J_1 C_2 J_2 > 0$ should be ensured (corresponding to identical Oersted contributions in both pillars). This excludes configurations Pc3 and APc3 from Tab. \ref{tab:tab1} from being optimal configurations for synchronization.

The fourth term in Eq. (\ref{eq:Thiele}) is the spin transfer force, which for the case of perpendicularly uniform magnetized polarizer writes\cite{Khvalkovskiy2009}:
\begin{equation}
\mv{F}^\text{STT}_i = \kappa\left( \mv{X}_i \times \muv{z} \right)
\end{equation}
where $\kappa = \pi \gamma a_J M_\text{s} h$ is the effective spin torque efficiency on the vortex and $a_J=\hbar p_z PJ/(2|e|hM_\text{s})$. In this study, we chose to neglect the field-like torque (FLT) contribution. While the FLT is negligible in case of a metallic intermediate layer, its amplitude can however reach a significant fraction of the Slonczewski torque in case of a magnetic tunnel junction. However, micromagnetic simulations computed with a FLT contribution of 10\% (typical) of the magnitude of the Slonczewski term showed no significant influence on the gyrotropic dynamics.
A last term accounts for the interaction dipolar force between the two neighbored vortices: $\mv{F}_\text{int,ji} (\mv{X}_{1,2}) = - \partial \MEAN{W_\text{int}}(\mv{X}_1,\mv{X}_2)/\partial \mv{X}_{1,2} = -C_1 C_2\me \mv{X}_{2,1}$, where $\me$ is either $\me^{P}$ or $\me^{AP}$ depending on $P_1.P_2$ sign.

The system of coupled equations for the vortices core motion given in Eq. (\ref{eq:Thiele}) provides a dynamical description of the phase locking between the two cores. We introduce the two variables $\Psi = P_1 \varphi_1 - P_2 \varphi_2$ and $\epsilon=(X_1-X_2)/(X_1+X_2)$. Following the methodology described it \citet{Belanovsky2012}, by linearizing the system around equilibrium trajectories, we obtain a linear set of equations describing the evolution in time of the relative phases and amplitudes:
\begin{subequations}
\begin{empheq}[left=\empheqlbrace]{align}
\dot{\varepsilon} &= -2\alpha \eta \left( \frac{\me}{G} + \omega_0 a r_{0}^2 \right) \varepsilon - \frac{\me}{G}\Psi \label{eq:eps_dot}\\
\dot{\Psi}        &= -4 \left( \frac{\me}{G} + \omega_0 a r_{0}^2 \right) \varepsilon + 2\alpha \eta \frac{\me}{G} \Psi \label{eq:psi_dot}
\end{empheq}
\end{subequations}
where $r_0 = X_0/R$ is the normalized average gyration radius and $a = k^\prime_\text{ms}/k_\text{ms} = 1/4$. The two equations (\ref{eq:eps_dot}) and (\ref{eq:psi_dot}) are linear and their eigenvalues are
\begin{equation*} 
\lambda_{1,2} = -\alpha \eta \omega_0 a r_{0}^2 \pm \sqrt{\alpha^2 \eta^2 \omega_0^2 (a r_{0}^2)^2 + 4 \frac{\me^2}{G^2} - 4 \frac{\me}{G} \omega_0 a r_{0}^2}.
\end{equation*}

In the case of periodic solutions, the phase-locking dynamics is characterized by a phase-locking time ($\tau$) and a beating frequency ($\Omega$) that can be written as:
\begin{subequations}
\begin{empheq}[left=\empheqlbrace]{align}
1/\tau   &= -\alpha \eta \omega_0 a r_{0}^2 \label{eq:tau}\\
\Omega^2 &= -(\alpha\eta\omega_0 a r_0^2)^2 - 4\left(\frac{\me}{G}\right)^2 - 4\frac{\me}{G}\omega_0 a r_{0}^2 \label{eq:Omega}
\end{empheq}
\end{subequations}

In next section, we propose to realize micromagnetic simulations\footnote{SpinPM is a micromagnetic code developed by the Istituto P.M. srl (Torino, Italy - www.istituto-pm.it) based on a forth order Runge-Kutta numerical scheme with an adaptative time-step control for the time integration}, from which $\Omega$ and $\tau$ will be extracted from the phase-locking dynamics. The effective coupling coefficient in each configuration $\me$ will then be derived for each considered configuration by simply reverting equations (\ref{eq:tau}) and (\ref{eq:Omega}):
\begin{equation}
\me(\tau,\Omega) = \frac{G}{2}\left(1/(\tau \alpha \eta) - \sqrt{1/(\tau \alpha \eta)^2-\Omega(L)^2}\right).
\label{eq:Wint_L}
\end{equation}

These micromagnetic simulations represent a more realistic picture of the coupled system as it takes into account the non-punctual geometry of the magnets as well as the full current induced Oersted field contribution, including cross-talk between nano-pillars.

\section{\label{sec:sim}Micromagnetic simulations}

\begin{figure}[!ht]
	\includegraphics[scale=1]{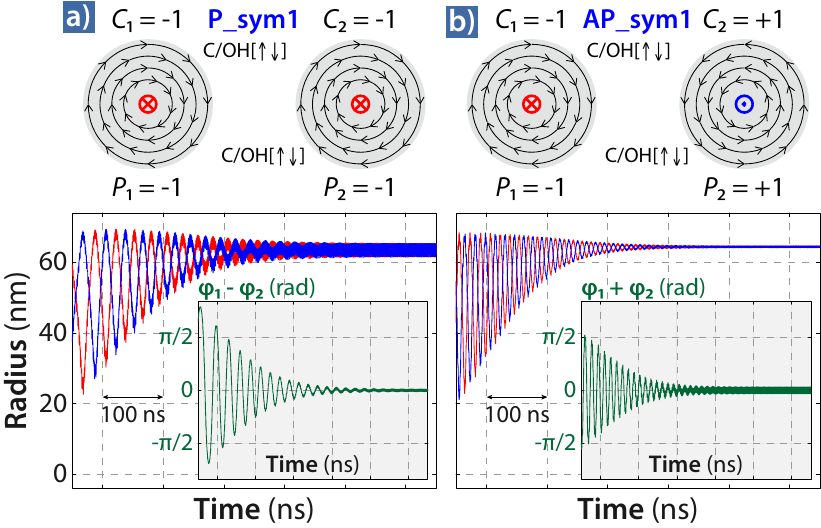}
	\caption{(Color online) a) (resp. b)) Vortex cores orbital radii and phase difference $\Psi = \varphi_1 - \varphi_2$ (resp. sum $\Psi = \varphi_1 + \varphi_2$) obtained by micromagnetic simulations for the Pc1 (resp. APc1) configuration where $L=50$ nm. As seen in top schematic illustrations the chirality (C) is opposite to the Oersted field (OH) in all the dots (C/OH[$\uparrow\downarrow$]).}
	\label{fig:sim1}
\end{figure}

We first compare the results of micromagnetic simulations obtained for the two cases Pc1 and APc1 with a separating distance between nano-pillars $L=50$ nm. The evolution of radii and dephasing parameter $\Psi$ is shown in figures \ref{fig:sim1}(a) and \ref{fig:sim1}(b) respectively and some numerical values are given in Tab. \ref{tab:tab2}. These results first confirm that phase-locking is achieved in both configurations. For both configurations, self-sustained unlocked oscillations in each pillar start at the same frequency but start with a random phase shift, and then converge towards phase-locked regime in very close phase-locking times ($\tau$). In their phase-locked state, both vortex cores oscillate with identical radii. 

\begin{table}[!ht]
	\centering
	\renewcommand{\arraystretch}{1.3}
	\begin{ruledtabular}
		\begin{tabular}{ccccc}
			config.  & $f$ (MHz) & $X_{01}$ (nm) & $X_{02}$ (nm) & $\Psi$ \\
			\hline
			Pc1  & 468.80 & 63.59 & 63.59 & $\rightarrow 0$ \\
			APc1 & 470.57 & 64.46 & 64.46 & $\rightarrow 0$ \\
			APc2 & 497.75 & 44.51 & 44.51 & $\rightarrow 0$ \\
			APc3 & 476.31 & 65.59 & 40.59 & $\rightarrow \pi$ \\
		\end{tabular}
	\end{ruledtabular}
	\caption{\label{tab:tab2}Numerical values of parameters extracted from micromagnetic simulations: $f$ is the common oscillation frequency, $X_{01}$ (resp. $X_{02}$) is the left (resp. right) dot vortex steady-state radius, and $\Psi$ is the dephasing parameter.}
\end{table}

\begin{table}[!ht]
	\centering
	\renewcommand{\arraystretch}{1.3}
	\begin{ruledtabular}
		\begin{tabular}{ccccc}
			config.  & \mytab{$\tau$}{(ns)} & \mytab{$\Omega$}{(MHz)} & \mytab{$\me/G$}{(MHz)} & \mytab{$\MEAN{ W_\text{int} }$}{($\times 10^{-14}$ erg)} \\
			\hline
			Pc1  & 82.78 & 40.136 & 19.7 & -22.75 [-27.08] \\
			APc1 & 71.20 & 67.380 & 49.2 & -58.31 [-64.23] \\
		\end{tabular}
	\end{ruledtabular}
	\caption{\label{tab:tab3}Numerical values of $\tau$, $\Omega$, $\me/G$, and $\MEAN{W_\text{int}}$ obtained after combining micromagnetic simulations and our Thiele equation approach for Pc1 and APc1 configurations. The last column containing the mean interaction energy computed by Eq. (\ref{eq:Wint_L}) shows also the numerical evaluation using Eq. (\ref{eq:Wnum}) inside brackets.}
\end{table}

The phase dynamics obtained by micromagnetic simulations are fitted to $\Psi = A e^{-t/\tau} \sin(\Omega t + \varphi_0)$ to extract $\Omega$ the beating frequency and $\tau$ the convergence time for phase-locking (see Tab. \ref{tab:tab3}). The effective coupling values for $L=50$ nm are then deduced: $\me/G=19.7$ MHz for Pc configuration, and $\me/G=49.2$ MHz for the APc one. The coupling strength then appears to be stronger in the AP configuration ($\sim 2.5 \times$) as expected from macro-dipole model.

\begin{figure}[!ht]
	\includegraphics[scale=1]{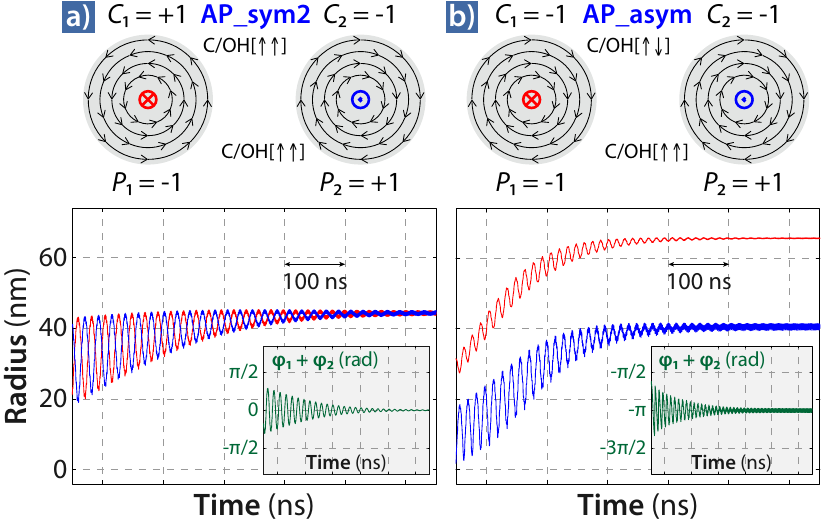}
	\caption{(Color online) Vortex cores orbital radii and phase sum $\Psi = \varphi_1 + \varphi_2$ where $L=50$ nm for configurations: a) "APc2" where both the vortex chiralities are aligned with the Oersted field (C/OH[$\uparrow\uparrow$]), b) "APc3" where the vortex in the right (resp. left) dot has an anti-aligned (aligned) chirality with the current induced Oersted field C/OH[$\uparrow\downarrow$] (resp. (C/OH[$\uparrow\uparrow$])).}
	\label{fig:sim2}
\end{figure}

The results for the "APc2" and the "APc3" configurations for $L=50$ nm are shown in Fig. \ref{fig:sim2}. Again in both cases phase-locking is achieved. In the symmetric case APc2, for which both chiralities are parallel to the Oersted field, starting frequencies are again identical in each pillar, whereas it is not the case for APc3 configuration, in which symmetry is broken by the Oersted-field being opposed to chirality in only one pillar. In the latter case, the two auto-oscillators have to adapt their frequencies to achieve synchronization to a common frequency $f_1 = f_2 = 476.31$ MHz, by shifting their amplitudes accordingly.
As highlighted previously, the micromagnetic simulations confirm that the equilibrium phase shift changes from $\ABS{\Psi} = 0$ to $\ABS{\Psi} = \pi$ when the sign of respective chiralities $\text{sign}(C_1 C_2)$ changes.

\section{Numerical approach}

To investigate further the difference in coupling strength between Pc and APc configurations, and validate the macro-dipoles approach, a more precise numerical calculation of the dipolar energy is proposed. The dipolar interaction energy is here summed up over the full magnetization distributions obtained by micromagnetic simulations. It consists in taking into account all the spin to spin i.e., cell to cell, interactions between the left pillar and the right pillar as follows:
\begin{equation}
W_\text{int}^\text{num} = \sum^{N_1}_{i=1} \sum^{N_2}_{j=1} W_{\text{int}, ij},
\label{eq:Wnum}
\end{equation}
where $W_{\text{int},ij} = -(3(\mv{\mu}_i\cdot\mv{e}_{ij})(\mv{\mu}_j \cdot \mv{e}_{ij}) - \mv{\mu}_i \cdot \mv{\mu}_j) / \Vert\mv{D}_{ij}\Vert^3$. $N_1$ (resp. $N_2$) is the number of cells in the left (resp. right) dot.

\begin{figure}[!ht]
	\includegraphics[scale=1]{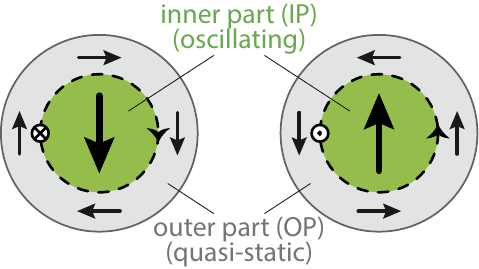}
	\caption{(Color online) Illustration of the in-plane magnetization of the two oscillating vortices. The gray zone represent the quasi-static magnetization (outer part - OP, $r>X_0$) and the green zone represents the oscillating magnetization (inner part - IP, $r<X_0$). The dashed lines give the vortex orbital movement delimitation where $r=X_0$.}
	\label{fig:ipop}
\end{figure}

\begin{figure}[!ht]
	\includegraphics[scale=1]{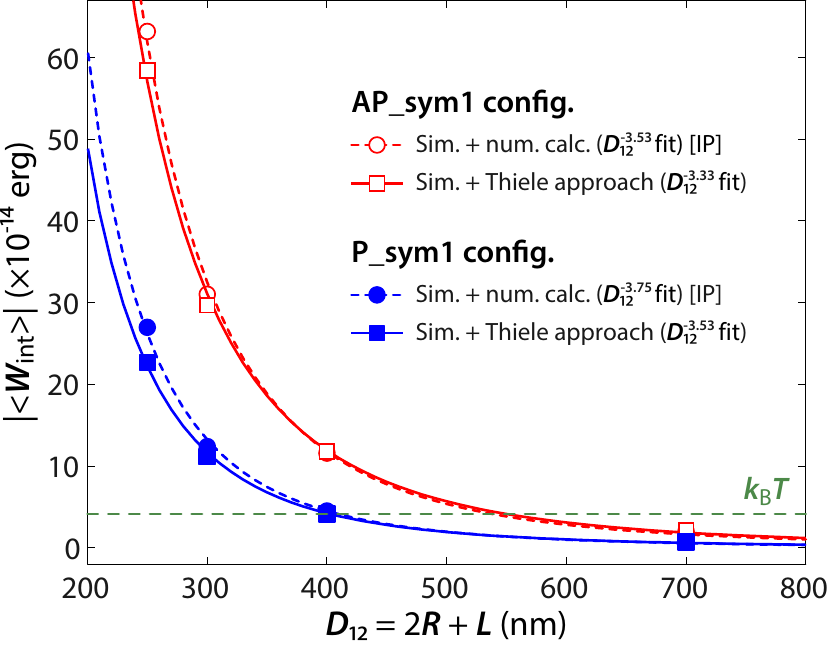}
\caption{(Color online) Inter-dot distance ($D_{12} = 2R + L$) dependence of the absolute value of the mean coupling energy $\MEAN{ W_\text{int} }$: The case for parallel (resp. anti-parallel) polarities using a macro-dipoles and Thiele equation approach model [filled squares (resp. unfilled squares)] and by numerical dipole-dipole computation of the inner parts (IP, see Fig. \ref{fig:ipop}) of the vortex cores trajectories [filled circles (resp. unfilled circles)] are shown in blue (resp. red).}
\label{fig:W1}
\end{figure}

As illustrated in Fig. \ref{fig:ipop} each dot can be seen as composed by two distinct regions. The outer part (OP) and the inner part (IP) with respect to the vortex gyrotropic trajectory. The OP is a quasi-static region and the inner part can be considered as an oscillating dipole. For means of comparison with our analytical model the OP region is first neglected. As seen in Fig. \ref{fig:W1} the values of $\MEAN{W^\text{num}_\text{int}}$ are close to the values of $\MEAN{W_\text{int}}$ obtained through the macro-dipoles and Thiele equation approach when the OP region is neglected.

\begin{figure}[!ht]
	\includegraphics[scale=1]{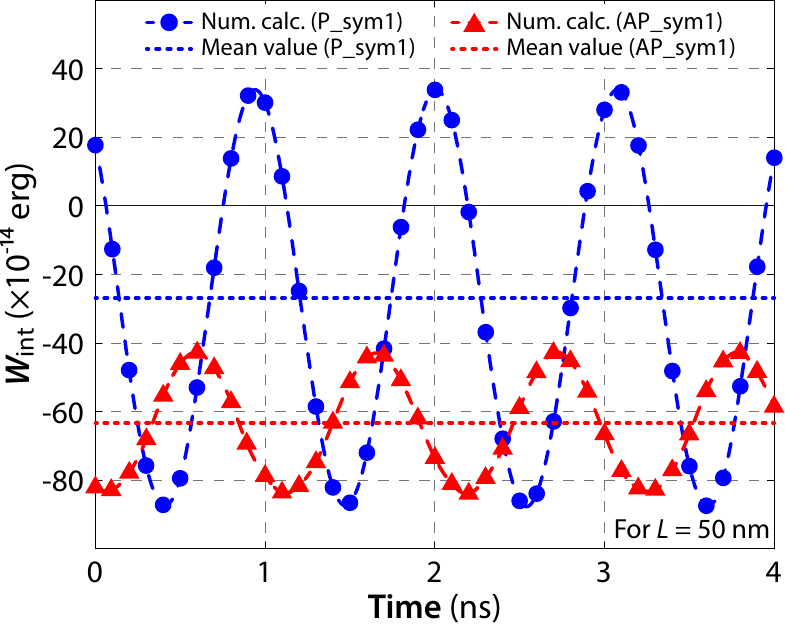}
	\caption{(Color online) Numerical computation of the dipolar energy ($W_\text{int}^\text{num}$) evolution of two interaction and synchronized vortices extracted from micromagnetic simulations for $L=50$ nm considering the IP region only (see Fig. \ref{fig:ipop}). The blue curve shows the evolution of the in phase oscillating vortices (with parallel polarities, P configuration) and the red one the anti-phase case (vortices with anti-parallel polarities, AP configuration). The dashed colored lines represent the corresponding mean values of the coupling energies ($\MEAN{ W_\text{int}^\text{num} }$).}
	\label{fig:Wosc}
\end{figure}

Figure \ref{fig:Wosc} shows the results for an edge to edge distance between two STVOs of $L=50$ nm considering the IP region only in which we compile the data for APc1 configuration (red triangles) and Pc1 configuration (blue dots). The dashed lines give the mean value of the interacting dipolar energy $\MEAN{ W_\text{int}^\text{num} }$ . In both cases and as expected, the energy $W_\text{int}^\text{num}$ oscillates at a frequency that corresponds to twice the gyrotropic frequency (see Tab. \ref{tab:tab2}).

We reproduced the process for several other distances between the dots ($L = {100,200,500}$ nm). The evolution of the average interaction energy versus $L$ extracted from micromagnetic simulations is shown in Fig. \ref{fig:W1}. For both Pc and APc configurations. The agreement between numerical and Thiele-based estimation of interacting energy is fairly good, notably for large interdot distance. 

As discussed theoretically in section \ref{sec:macrodip}, the macro-dipole model gives a ratio of 3 between the interaction energy for the P and AP core configurations. In contrast to that prediction, it should be noticed that for small values of the edge to edge interdot distance $L$ the ratio between the calculated energy gets smaller ($\sim$2.6 for $L=50$ nm). 

\begin{figure}[!ht]
\includegraphics[scale=1]{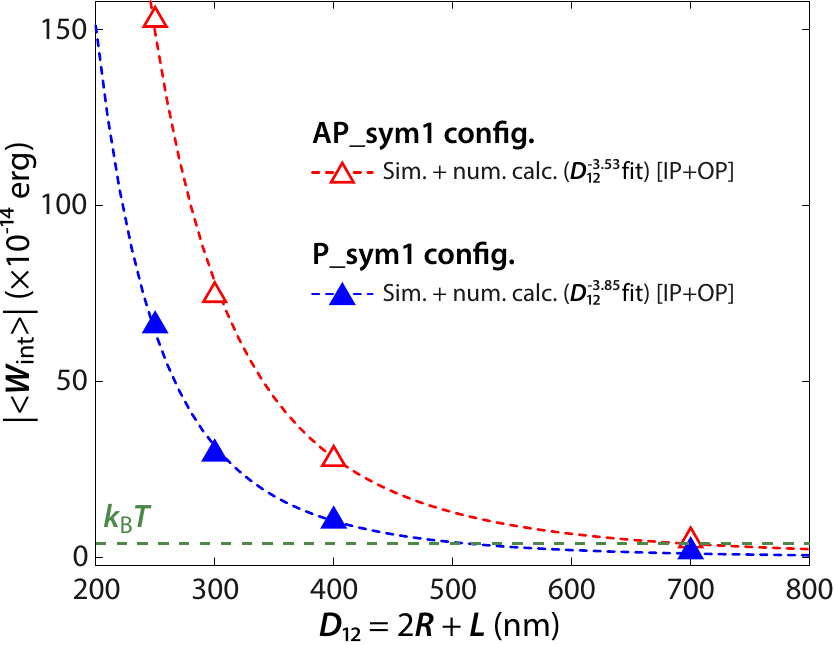}
\caption{(Color online) Inter-dot distance ($D_{12} = 2R + L$) dependence of the absolute value of the coupling energy ($W_\text{int}$): The case for parallel (Pc) [resp. anti-parallel (APc)] polarities obtained by numerical dipole-dipole computation for the whole volume (IP + OP) [filled triangles (resp. unfilled triangles)] are shown in blue (resp. red).}
\label{fig:W2}
\end{figure}

The thermal fluctuations were not involved in our simulations. Yet, the obtained mean coupling energy $\ABSMEAN{W_\text{int}}$ can be easily compared against thermal energy ($k_\text{B}T$). The condition for stable synchronization i.e., $\ABSMEAN{W_\text{int}} > k_\text{B}T$, are then found to be $D_{12} < 400$ nm in the Pc1 configuration and $D_{12} < 550$ nm in the APc1 configuration. Finally, the data is fitted with a decay law for $W_\text{int}$ as function of $D_{12}$, i.e. $D_{12}^{-\sigma}$ where $\sigma$ has values between 3 and 4 as shown in of figures \ref{fig:W1} and \ref{fig:W2}.

\section{Conclusion}

In conclusion, we performed a comparative study of vortices parameters configuration for the synchronization of two dipolarly coupled spin transfer vortex-based oscillators. As the major result of this numerical and analytical study, we demonstrate that the effective coupling of two vortices with opposite core polarities and hence gyrating in opposite directions is larger than the case of identical polarities. 

By studying different contributions to the coupled vortices dynamics, we have also shown that this configuration matches with feasible experimental configuration. Optimal configuration then corresponds to nano-pillars connected in series.

Comparing the computed $W_\text{int}$ (IP only) with the thermal energy $k_\text{B}T$, one obtains that synchronization can be presumably achieved $D_{12} \leqslant 400$ nm in case of parallel polarities configuration (Pc), while $D_{12} \leqslant 550$ would be sufficient in case of anti-parellel polarities.

As far as phase-locking stability is concerned, we highlighted that the dipolar interaction keeps involving strong oscillations in the coupling energy even after achieving synchronization. These interactions will play against synchronization and should decrease the minimum inter-pillar distance to achieve synchronization.

\section{Acknowledgments}

F.A.A. acknowledges the Research Science Foundation of Belgium (FRS-FNRS) for financial support (FRIA grant). The authors acknowledge also the ANR agency (SPINNOVA ANR-11-NANO-0016) and the EU FP7 grant (MOSAIC No. ICT-FP7- n.317950) for financial support. This publication is based on work funded by Skolkovo Institute of Science and Technology (Skoltech) within the framework of the Skoltech/MIT Initiative.

\bibliography{../MyLib.bib}

\end{document}